\documentclass[12pt,preprint]{aastex}
\usepackage{graphicx}

\newcommand {\ea} {{\it et~al.}}
\newcommand {\be} {\begin{equation}}
\newcommand {\ee} {\end{equation}}

\shorttitle{Emission Models of Luminous Blazars}

\shortauthors{Sikora \ea}

\begin{document}

\title{Constraining Emission Models of Luminous Blazar Sources}

\author{Marek~Sikora\altaffilmark{1},
  {\L}ukasz~Stawarz\altaffilmark{2,3,4},
  Rafa{\l}~Moderski\altaffilmark{1},
  Krzysztof~Nalewajko\altaffilmark{1}, and
  Greg~M.~Madejski\altaffilmark{2,4}}

\altaffiltext{1}{Nicolaus Copernicus Astronomical Center, Bartycka 18,
  00-716 Warsaw, Poland; \tt{sikora@camk.edu.pl}}
\altaffiltext{2}{Kavli Institute for Particle Astrophysics and
  Cosmology, Stanford University, Stanford, CA 94305, USA}
\altaffiltext{3}{Astronomical Observatory, Jagiellonian University,
  ul. Orla 171, 30-244 Krak\'ow, Poland} \altaffiltext{4}{Stanford
  Linear Accelerator Center, 2575 Sand Hill Road, Menlo Park, CA
  94025, USA}

\begin{abstract}

Many luminous blazars which are associated with quasar-type active
galactic nuclei display broad-band spectra characterized
by a large luminosity ratio of their
high-energy ($\gamma$-ray) and low-energy (synchrotron) spectral
components.  This large ratio, reaching values up to $100$, 
challenges the standard synchrotron self-Compton models by means
of substantial departures from the minimum power condition.
Luminous blazars have also typically very hard X-ray
spectra, and those in turn seem to challenge
hadronic scenarios for the high energy blazar emission. 
As shown in this paper, no such problems are faced by
the models which involve Comptonization of radiation provided
by  a broad-line-region, or dusty molecular torus.  
The lack or weakness of bulk Compton 
and Klein-Nishina features indicated by the presently available data 
favors production of $\gamma$-rays via 
up-scattering of infrared photons from hot dust.
This implies that  the blazar emission zone is located at parsec-scale 
distances from the nucleus, and as such  is possibly associated with
the extended, quasi-stationary reconfinement shocks formed in relativistic 
outflows. This scenario predicts characteristic timescales for flux changes
in luminous blazars to be days/weeks, consistent with the  
variability patterns observed in such systems 
 at infrared, optical and $\gamma$-ray frequencies.  
We also propose that the parsec-scale
blazar activity can be occasionally accompanied by dissipative events 
taking place at sub-parsec distances and powered by internal shocks and/or
reconnection of magnetic fields.  These could account for the
multiwavelength intra-day flares occasionally observed in powerful
blazars sources. 

\end{abstract}

\keywords{galaxies: active --- galaxies: jets --- gamma-rays: observations --- quasars: general --- radiation mechanisms: non-thermal --- acceleration of particles}

\section{Introduction}

Blazars are active galactic nuclei (AGN) dominated by Doppler
boosted radiation of relativistic jets, and as such, they provide an
exceptional opportunity for studying the physics of innermost portions
of AGN jets. However, in order to take advantage of this opportunity,
the appropriate radiative processes responsible for generation of a
broad-band jet emission must be identified.  And, while the low-energy
component of double-peaked blazar spectra is uniquely identified as
originating via the synchrotron radiation of ultrarelativistic
electrons (hereafter by electrons we mean both electrons and
positrons), the origin of the high-energy spectral component is still
under debate. Possible contributions include inverse-Compton (IC) emission
by electrons directly accelerated within the jet (and producing the
observed synchrotron continuum), synchrotron radiation of pair
cascades powered by hadronic processes, and synchrotron emission of
ultra-high-energy protons and muons (see reviews by Sikora \& Madejski 
2001; Levinson 2006; B\"ottcher 2007). Seed photons for the IC process
may be provided by the jet synchrotron emission as well as by the
external photon sources, such as an accretion disk, 
broad line regions (BLR) and/or dusty
tori (hereafter `hot dust region', HDR). The
IC models involving Comptonization of local synchrotron radiation are
called synchrotron-self-Compton (SSC) models, and those involving
Comptonization of external radiation -- external-radiation-Compton,
or, for short, external-Compton (EC) models.

In this paper we focus on blazars with a dense radiative circumnuclear
environment, i.e. those hosted by quasars.  High-energy (HE;
$\gamma$-ray) luminosities of many of those objects
exceed their low-energy (LE; synchrotron) luminosities by a factor $q
\equiv L_{HE}/L_{LE} = 10-100$.  At the same time, their X-ray spectra
are often very hard, in a number of cases characterized by energy spectral
indices $\alpha_x < 0.5$ (see Table 1 and references therein).
Constraints resulting from such properties are discussed
in \S\,2 and \S\,3, respectively. In addition,
we investigate constraints imposed: by the spectral localization of the
two prominent blazar radiative components (\S\,4); by the lack of
theoretically predicted bulk-Compton features in blazar spectra
(\S\,5); and by the lack of signatures of the
Klein-Nishina  features in their synchrotron continua (\S\,6).  We
discuss EC(BLR) and EC(HDR) models in \S\,7, highlighting 
their similarities and differences.
We discuss the possibility of detecting electromagnetic imprints of the
presence of protons in the blazar jet spectra in \S\,8. Our final 
results regarding physics of the parsec-scale jets of quasars
are presented in \S\,9, and we briefly conclude in \S\,{10}.
Throughout the paper the approximation $\theta_j = \theta_{obs} =
1/\Gamma$ is used, where $\theta_j$ is the opening angle of a conical
jet, $\theta_{obs}$ is the angle between the line of sight and the jet
axis, and $\Gamma$ is the jet Lorentz factor.  We also neglect
the redshift dependence in the derived expressions, but note that 
this dependence goes at most as a factor $(1+z)$.  

\section{Large Luminosity Ratio}

Since our understanding of particle acceleration processes in
relativistic plasma is limited (and relies primarily on results
obtained strictly in the test-particle limit), the existing hadronic
models are not sufficiently quantitative to make robust predictions
regarding the relative power required to be injected in the form of
high energy protons and ultrarelativistic electrons.  Hence, the
observed large luminosity ratio between the high- and the low-energy
components in blazar spectra can be scaled by a free parameter in
those models.  In contrast, in leptonic scenarios both components are
produced by the same population of ultrarelativistic electrons, and
therefore the observed large values of $q$ provide valuable
constraints on the blazar parameters.

Noting that for luminous, quasar-associated blazars, the
Klein-Nishina and the pair production effects
are likely to be insignificant (Moderski et al. 2005; see also \S\,6),
and that the effect of the anisotropy of external-Compton scattering
in the source co-moving frame is negligible for $\theta_{obs} \sim
1/\Gamma$ (Dermer 1995; Moderski et al. 2003), the following ratios of
energy densities can be approximated as being equal:
\be {u_{SSC}' \over u_{syn}'} \simeq {u_{syn}' \over u_B'} \simeq
    {u_{EC}' \over u_{ext}'} \equiv A \, , \ee
where $A$ is the `Thomson amplification factor', $u_B'$ is the
comoving magnetic energy density, while $u_{syn}'$, $u_{SSC}'$,
$u_{ext}'$ and $u_{EC}'$ are the energy densities of the synchrotron,
SSC, external (BLR or HDR), and EC photon fields, respectively, all as
measured in the source rest frame.

\subsection{Large Luminosity Ratio in the SSC Model}

In the case of the SSC emission dominating the high-energy spectral
component, $L_{HE} \simeq L_{SSC}$, one has
\be q \simeq {L_{SSC} \over L_{syn}} \simeq {u_{SSC}' \over u_{syn}'}
\simeq {u_{syn}' \over u_B'} = A \, , \ee
and
\be {u_B' \over u_{SSC}'} \simeq {u_B' \over u_{syn}'} \, {u_{syn}'
  \over u_{SSC}'} = A^{-2} = q^{-2} \, , \ee
where $L_{syn}$ and $L_{SSC}$ are the observed synchrotron and SSC
luminosities, respectively.  Noting next that the approximate relation
between the ratio of total energy densities stored in non-thermal
photons, $u_{rad}'$, and electrons, $u_e'$, is
\be 
{u_{rad}' \over u_e'} \simeq {\eta_{rad} \over 1-\eta_{rad}} \, , 
\ee
where $\eta_{rad}$ is the radiative efficiency of the injected
electrons, and that in the SSC model with $q \gg 1$ one has $u_{rad}'
\simeq u_{SSC}'$, we obtain
\be 
{u_B' \over u_e'} \simeq {u_B' \over u_{SSC}'} \, {u_{SSC}' \over u_e'} \simeq 
{1 \over q^2} \, {\eta_{rad} \over (1-\eta_{rad})} \, .
\ee
Since for the luminous blazars $\eta_{rad} > 0.5$, the `high-$q$
states' of the SSC-dominated $\gamma$-ray emission can be reached only
for $u_e' / u_B' > q^2 \gg 1$, i.e. for conditions
far away from equipartition.

\subsection{Large Luminosity Ratio in the EC Models}

In the case of the high-energy emission dominated by the EC radiation
we have $L_{HE} \simeq L_{EC}$, leading to 
\be q = {L_{EC} \over L_{syn}} \simeq {u_{EC}' \over u_{syn}'} \simeq
    {u_{ext}' \over u_B'} \, , \ee
and 
\be 
{u_B' \over u_{EC}'} \simeq {u_B' \over u_{syn}'} \, {u_{syn}' \over u_{EC}'} = A^{-1} \, q^{-1} \, . 
\ee
Hence, with the further approximation $u_{rad}' \simeq u_{EC}'$ (as
justified for $q \gg 1$), we obtain
\be 
{u_B' \over u_e'} \simeq {u_B' \over u_{EC}'} \, {u_{EC}' \over u_e'} \simeq 
{1 \over A \, q} \, {\eta_{rad} \over (1-\eta_{rad})} \, .
\label{ubue}
\ee
Obviously, in the EC models the quantities $A$ and $q$ are not
equivalent, and the equipartition between magnetic field and electron
energy densities can be reached for any (large) value of the $q$
parameter provided
\be 
A = {1 \over q} \, { \eta_{rad} \over (1-\eta_{rad})} \, . 
\ee
Moreover, since $A \simeq u_{EC}' / u_{ext}'$, where
\be 
u_{EC}' \simeq {L_{EC} \over 4 \pi R^2 c \Gamma^4}
\ee
and
\be 
u_{ext}' \simeq {\Gamma^2 \xi_{ext} L_{disk} \over 4 \pi r^2 c} \, , 
\ee
one may find that
\be 
A = {L_{EC} \over \xi_{ext} L_{disk} \Gamma^4}  \, .
\ee
In the above, we have introduced the distance of a blazar emission
zone from the nucleus $r$, the radius of the emission zone $R
\simeq r \, \theta_j \simeq r/\Gamma$, the accretion disk luminosity
$L_{disc}$, and the fraction $\xi_{ext}$ of the disk radiation
reprocessed in the BLR or HDR. This, when combined with
Eq.~(\ref{ubue}), shows that the minimum power condition $u_B' \sim
u_e'$ implies bulk Lorentz factors of the emitting plasma
\be \Gamma \simeq 20 \times \left[\left({q \over 10}\right)
  \left({\xi_{ext} \over 0.1}\right)^{-1} \left({L_{EC} \over
    10^{49}\,{\rm erg\,s^{-1}}}\right) \left({L_{disk} \over
    10^{46}\,{\rm erg\,s^{-1}}}\right)^{-1} \left( {1 -\eta_{rad}
    \over \eta_{rad}} \right)\right]^{1/4} \, .
\label{doppler} 
\ee
We note that typically in blazar sources the disk luminosity
$L_{disk}$ cannot be observed directly but instead can be estimated
from the observed luminosity of broad emission lines, leading to
$L_{disk} \gtrsim 10^{46}$\,erg\,s$^{-1}$.  Also, the derived
`equipartition' value of the jet Lorentz factor, which depends only
weakly on the model parameters, is nicely consistent with the typical
values implied by the observed superluminal expansion of pc/kpc-scale
radio blobs in $\gamma$-ray bright, quasar-hosted blazars (Jorstad et
al. 2001a, 2001b; Lister et al. 2009; Kovalev et al. 2009).

\section{Hard X-ray Spectra}

About $30\%$ of blazars detected by BeppoSAX up to the energy of
$50$\,keV are characterized by very hard X-ray spectra, with the X-ray
spectral indices smaller than the `canonical' value $0.5$ (Giommi et
al. 2002; Donato et al. 2005; see also the Table~1).  In several cases
strong indications were found for such hard spectra to extend
up to at least $\sim 100$\,keV.  Such hard spectra must be produced in
a slow cooling regime for the radiating particles, and below we
investigate what constraints such observations impose on different
emission models.

\subsection{Hard X-ray Spectra in the Hadronic Models}

\subsubsection{Proton-Initiated Cascades}

In the hadronic models for luminous blazars, the observed X-ray
emission is produced by synchrotron radiation of pair cascades powered
by ultrarelativistic protons (Mannheim \& Biermann 1992; Mannheim
1993).  In order to obtain $\alpha_x < 0.5$, the cooling break in the
electron distribution formed by such cascades, corresponding to the
cooling break frequency in the synchrotron continuum $\nu_{syn,\,c}$
as measured in the observer frame, must correspond to the electron Lorentz
factor
\be \gamma_c \simeq 6 \times 10^5 \, {B_{\rm G}'}^{-1/2} \, \left( h
\nu_{syn,\,c} \over 100 \, {\rm keV} \right)^{1/2}\, \left(\Gamma
\over 20\right)^{-1/2} \, , \ee
where the comoving magnetic intensity is expressed in units of
Gauss, $B_{\rm G}' \equiv B'/{\rm G}$.  Meanwhile, equating the time
scale of the synchrotron (electron) energy losses $t_{cool}' \simeq
\gamma/|\dot \gamma_{syn}|$ with the dynamical time scale of the
emission region $t_{dyn}' \simeq (R/c) \simeq r/(c\Gamma)$, one may
find
\be 
\gamma_c \simeq {m_e c^2 \, \Gamma \over \sigma_T r u_B' } \, .
\ee
Hence, hard X-ray spectra observed in luminous blazars, if powered by
hadronic interactions, require jet magnetic field as low as
\be B_{\rm G}' \simeq 5 \times 10^{-3} \, \left(\Gamma \over 20\right)
\, \left(r \over {\rm pc}\right)^{-2/3} \left(h\nu_{syn, \, c} \over
100 \, {\rm keV}\right)^{-1/3} \, .  
\label{bgp}
\ee
However, in such weak magnetic fields, the SSC spectral component
produced by copious, directly accelerated electrons can lead to
overproduction of the X-ray flux.  In order to avoid this situation,
the blazar emission zone must be located at sufficiently large
distances to satisfy $A = u_{syn}' / u_B' < q$, where the comoving
energy density of the the jet photons is
\be u_{syn}' \simeq {L_{syn} \over 4 \pi r^2 c \Gamma^2} \simeq 7
\times 10^{-5} \, \left(L_{syn} \over 10^{47} \, {\rm erg \, s^{-1}}
\right) \, \left( r \over {\rm pc} \right)^{-2} \, \left( \Gamma \over
20 \right)^{-2} \,\, {\rm erg \, cm^{-3}} \, .  \ee
This condition, together with Eq.~(\ref{bgp}), implies in turn
\be r > 20 \, \left(L_{syn} \over 10^{47} \, {\rm erg \,
  s^{-1}}\right)^{3/2} \left (h \nu_{syn,\,c} \over 100 \, {\rm
  keV}\right) \, \left(\Gamma \over 20\right)^{-6} \, \left(q \over
10\right)^{-3/2} \,\, {\rm pc} \, .  
\ee

At such distances, however, pair cascades are likely to be produced very 
inefficiently. In order to verify this, we calculate below the efficiency 
of the photo-meson process which dominates the powering of the cascades. 
The opacity for the photo-meson process involving protons with random 
Lorentz factor $\gamma_p$ is
\be 
\tau_{p\gamma}(\gamma_p) = {{t'}_{p\gamma}^{-1} \over {t'}_{dyn}^{-1}} \simeq
\langle\sigma_{p\gamma} K_{p\gamma}\rangle \, {r \, n_{ph}'\!(\gamma_p) \over \Gamma} \, ,
\ee
where $\langle\sigma_{p\gamma} K_{p\gamma}\rangle \simeq 0.7 \times
10^{-28}$\,cm$^2$ is the product of the photo-meson cross-section and
inelasticity parameter averaged over the resonant energy range (Begelman, 
Rudak, \& Sikora 1990), $n_{ph}'(\gamma_p) =
\int_{\nu_{th}'(\gamma_p)} {n_{ph,\,\nu'}' d\nu'} $ is the number
density of photons with energies above the photo-meson threshold $h
\nu_{th}'(\gamma_p) \simeq m_{\pi} c^2 /\gamma_p$, and $m_{\pi} c^2
\simeq 140$\,MeV is the rest-mass energy of a pion.  For
$\tau_{p\gamma}(\gamma_p) < 1$ (which can be verified {\it a
  posteriori}), the total efficiency of the process is
\be
\label{eta_pgamma}
\eta_{p\gamma} = 
{\int_1^{\gamma_{p,\,max}} {\tau_{p\gamma} n_{\gamma_p}' \gamma_p \, d\gamma_p} \over
\int_1^{\gamma_{p,\,max}} { n_{\gamma_p}' \gamma_p \, d\gamma_p}} \, , 
\ee
where $n_{\gamma_p}'$ is the energy distribution of relativistic
protons.  The maximal available proton Lorentz factor, in turn,
$\gamma_{p,\,max}$, may be evaluated very roughly through the
condition $t_{acc}'(\gamma_p) = t_{dyn}'$ (since the dynamical
time scale is expected to be much shorter than any proton cooling
time scale), where
\be
\label{t_acc}
t_{acc}'(\gamma_p) \simeq {f R_{L} \over c} \simeq 
10^{-4} \gamma_p \, f \, {B_{\rm G}'}^{-1} \,\, {\rm s}
\ee
is the proton acceleration timescale, $R_{L} \simeq \gamma_p m_p c^2
/(e B')$ is the proton Larmor radius, and the factor $f$ accounts for
the specific particle acceleration model assumed, being generally
expected to be $f \geq 10$ (see Aharonian et al. 2002). 
This, together with Eq.~(\ref{bgp}), gives
\be \gamma_{p,\,max} \simeq 5 \times 10^{9} \, B_{\rm G}' \, \left({f
  \over 10}\right)^{-1} \left(r \over {\rm pc}\right) \, \left(\Gamma
\over 20\right)^{-1} \simeq 8 \times 10^7 \, \left({f \over
  10}\right)^{-1} \left(r \over 20 \, {\rm pc}\right)^{1/3} \left(h
\nu_{syn,\,c} \over 100 \, {\rm keV}\right)^{-1/3} \, ,  \ee
and the corresponding photon threshold energy is
$h \nu_{th}'(\gamma_{p,\,max}) \sim (\gamma_{p,\,max}/10^8)^{-1}$\,eV.

The target soft photons for the photo-meson production are provided by
the synchrotron radiation of directly accelerated jet electrons, as
well as by the external photon fields, dominated at parsec distances
by hot/warm dust located in the surrounding molecular tori.
The number density of synchrotron photons with energies larger than 
$h \nu_{th}(\gamma_{p,\,max})$ is
\begin{eqnarray}
& & n_{syn}'(\gamma_{p,\,max}) \simeq {L_{syn} \zeta(\gamma_{p,\,max}) \over 4 \pi r^2 c \, \Gamma^2 \,
h \nu_{th}'\!(\gamma_{p,\,max}) } \simeq \nonumber \\
&& 5 \times 10^4 \, \zeta(\gamma_{p,max}) \, \left({f \over 10}\right)^{-1} \left(L_{syn} \over 
10^{47} \, {\rm erg \, s^{-1}}\right)
\left(r \over 20 \, {\rm pc}\right)^{-5/3} \left(\Gamma \over 20\right)^{-2}  
\left(h \nu_{syn, \, c} \over 100\, {\rm keV}\right)^{1/3} \, {\rm cm^{-3}} \, , 
\end{eqnarray}
where $\zeta(\gamma_{p,\,max}) \simeq L_{syn}(\nu > \Gamma\,
\nu_{th}'(\gamma_{p,\,max}))/L_{syn}$ which for the typical synchrotron spectra
of luminous blazars and $\gamma_{p,\,max} \lesssim 10^8$ is expected to be 
$\ll 1$. 

As for the external target photon field, we note that the dust located
at $r \sim 20$\,pc and heated by the accretion disk with luminosity
$L_{disk} \sim 10^{46}$\,erg\,s$^{-1}$ has a temperature $T \sim
400$\,K (see Eq.~\ref{r_hdr} below).  Its electromagnetic spectrum
peaks at $h \nu_{ext} \simeq 0.1$\,eV, and its comoving number density
is
\be 
n_{ext}' \simeq \Gamma \, n_{ext} \simeq 
9 \times 10^7 \, \left( \xi_{ext} \over 0.1 \right) \, 
\left( L_{disk} \over 10^{46}\,{\rm erg \, s^{-1}} \right) \, 
\left( \Gamma \over 20 \right) \, \left( r \over 20\,{\rm pc} \right)^{-2} 
\left(h \nu_{ext} \over 0.1\,{\rm eV}\right)^{-1} \,\, {\rm cm^{-3}} 
\ee
(see Eq.~\ref{u_hdr} below).  Hence, the target radiation field for
the photo-meson process is dominated by the external radiation, and
thus the opacity for protons with energies $\gamma_p > \gamma_{p,\,th}
\simeq m_{\pi}{\rm c}^2 /(\Gamma\, h\nu_{ext})$ can be written as
\be \tau_{p\gamma}^{(ext)} \simeq 2 \times 10^{-2} \, \left(\xi_{ext}
\over 0.1 \right) \, \left( L_{disk} \over 10^{46}\,{\rm erg \,
  s^{-1}} \right) \left( r \over 20 \,{\rm pc} \right)^{-1}
\left(h\nu_{ext} \over 0.1\,{\rm eV}\right)^{-1} \, .  
\ee
For the most optimistic acceleration scenario with $f=10$, and for the
proton energy distribution of the form $n_{\gamma_p}' \propto
\gamma_p^{-2}$, this gives the total efficiency
\be \eta_{p\gamma}^{(ext)} \sim \tau_{p\gamma}^{(ext)}
    {\ln{(\gamma_{p,\,max}/\gamma_{p,\,th})} \over \ln{\gamma_{p,\,max}}}
    \sim 2 \times 10^{-4} \, . 
\ee
With such a low efficiency of the photo-meson production process, the jet power 
required to explain the observed $\gamma$-ray luminosities of the order of 
$10^{48}$\,erg\,s$^{-1}$ is
\be L_j \simeq {L_{\gamma} \over \Gamma^2 \, \eta_{diss} \eta_{p}
  \eta_{p\gamma}} \simeq 5 \times 10^{50} \, \left(L_{\gamma} \over
10^{48}\, {\rm erg \, s^{-1}} \right) \, \left(\eta_{p\gamma} \over
10^{-4}\right)^{-1} \left(\eta_{diss} \over 0.1 \right)^{-1}
\left({\eta_p \over 0.5}\right)^{-1} \,\, {\rm erg\,s^{-1}} \, , \ee
where $\eta_{diss}$ is the fraction of the jet energy flux dissipated
in the blazar emission zone, and $\eta_p$ is the fraction of the
dissipated energy channeled into relativistic protons. The required
jet kinetic luminosity is therefore \emph{orders of magnitude} larger
than the Eddington luminosity of a black hole with mass $10^9 \,
M_{\odot}$. Such serious energetic difficulties could
be overcome only by postulating \emph{ad hoc} some substructure of the blazar 
emission zone, involving spatial separation, and very different physical 
conditions characterizing regions where hadronic and leptonic processes 
operate.  However, several arbitrary assumptions involved in such a scenario 
as well as an increased number of free parameters would make this 
much less attractive.

\subsubsection{Proton-Synchrotron Emission}

Alternatively, in order to explain hard X-ray spectra of luminous
blazars still in the framework of hadronic models, one may postulate
that the entire observed high-energy (X-ray-to-$\gamma$-ray) flux is
dominated by the direct synchrotron radiation of ultrarelativistic
protons.  Noting the relation between the characteristic proton- and
electron-related synchrotron frequencies
$\nu_{p,\,syn}(\gamma_p=\gamma_e) = (m_e/m_p) \, \nu_{e,\,syn}$, and
the relation between the corresponding cooling rates $|\dot
\gamma_p|_{syn}(\gamma_p=\gamma_e) = (m_e/m_p)^3 \, |\dot
\gamma_e|_{syn}$ (see, e.g., Aharonian 2000), one can find the
appropriate radiative efficiency around the $\gamma$-ray luminosity
peak
\be 
\tau_{p,\,syn} (\gamma_{p,\,peak}) \equiv {{t'}_{p,\,syn}^{-1} \over {t'}_{dyn}^{-1}}
\simeq 2 \times 10^{-4} \, {B_{\rm G}'}^{3/2} \,
\left(r \over {\rm pc} \right) \, \left( \Gamma \over 20 \right)^{-3/2} \,  
\left(h\nu_{peak} \over 10 \,{\rm MeV} \right)^{1/2} \, ,
\ee
with
\be 
\gamma_{p,\,peak} \simeq 3 \times 10^8 \, {B_{\rm G}'}^{-1/2} \,
\left(h \nu_{peak} \over 10\,{\rm MeV} \right)^{1/2} \, . 
\ee
The resulting efficiency of the synchrotron proton emission is, again,
very low, and can be increased only by assuming very large magnetic
field intensity within the emission zone.  However, since the magnetic
energy flux is limited by the total jet kinetic power, $L_B \simeq c
u_B' \pi r^2 < L_j$, magnetic fields $B_{\rm G}' \gg 1$ are expected
only at $r \ll 1$\,pc.  In this context, one has to keep in mind that
at small distances from the black hole, AGN jets are still in the
acceleration phase, and thus the radiation produced there is not
boosted sufficiently strongly to account for the observed large
luminosities. The distance scale of the jet acceleration, which corresponds
roughly with the distance of the conversion of the Poynting flux dominated jet 
to that dominated by matter, is expected to be 
$r_0 \ge 10^3 R_g \simeq 0.03\,(M_{BH}/10^9\,M_{\odot})$ pc 
(Komissarov et al. 2007), where $R_{g}$ is the radius of the 
central black hole. Considering those additional constraints, we 
obtain 
\be \tau_{p,\,syn}(\gamma_{p,\,peak}) < 2 \times 10^{-3} \, \left(r_0
\over 0.03\,{\rm pc} \right)^{-1/2} \left(\Gamma \over
20\right)^{-3/2} \, \left(h \nu_{peak} \over 10 \,{\rm MeV}
\right)^{1/2} \, \left( L_j \over 10^{47}{\rm
  erg\,s^{-1}}\right)^{3/4} \, , 
\ee
where $L_j \sim 10^{47}$\,erg\,s$^{-1}$ corresponds to the Eddington
luminosity of the $10^9\,M_{\odot}$ black hole.  With such an
efficiency, the proton synchrotron peak luminosity is
\be L_{p,\, syn}(10\,{\rm MeV}) < 4 \times 10^{45} \,
\left(\eta_{diss} \over 0.1\right) \, \left(\eta_p \over 0.5 \right)
\, \left( \tau_{p,\, syn} \over 2 \times 10^{-3}\right) \,
\left(\Gamma \over 20\right)^2 \, \left(L_j \over 10^{47}\, {\rm erg
  \, s^{-1}} \right) \,\, {\rm erg \, s^{-1}} \, , 
\ee
i.e. orders of magnitude below the observed $\gamma$-ray luminosity
of powerful blazars.

\subsection{Hard X-ray Spectra in the SSC Model}

In the SSC model for large-$q$ blazars, the electron cooling rate is 
dominated by Comptonization of co-spatially produced synchrotron radiation, 
and therefore reads as
\be 
|\dot \gamma| \simeq |\dot \gamma|_{SSC} \simeq 
{c \sigma_T \gamma^2 u_{syn}' \over m_e c^2} \, , \ee
where 
\be 
u_{syn}' \simeq {L_{syn} \over 4 \pi R^2 c\Gamma^4} \, . 
\ee 
Balancing next the corresponding comoving time scale of radiative
losses $t_{cool}' \simeq \gamma/|\dot \gamma|_{SSC}$ with the
dynamical time scale of the emission region $t_{dyn}'$, one may find
the critical electron energy corresponding to the transition between
fast- and slow-cooling regimes,
\be \gamma_c \simeq 10^5 \, \left({\Gamma \over 20}\right)^3 \left({r
  \over {\rm pc}}\right) \left({L_{syn} \over 10^{47}\,{\rm
    erg\,s^{-1}}}\right)^{-1} \, .  
\ee
Electrons with $\gamma < \gamma_c$ cool inefficiently,
Compton-upscattering the far-infrared synchrotron radiation up to MeV
photon energies. Therefore, the observed hard X-ray spectra 
can be reproduced by the SSC model in a framework of 
the standard approach involving injection
of some given (single or broken) power-law electron energy distribution to the
emission region and its subsequent radiative cooling, provided the low-energy
segment of the injected electron energy distribution
is not softer that $\propto \gamma^{-2}$.

\subsection{Hard X-ray Spectra in the EC Models} 

In the EC models for blazars with large values of the $q$ parameter,
radiative cooling of electrons is dominated by Comptonization of the
external radiation fields.  As before, by equating the appropriate
cooling time scale
\be 
t_{cool}' \simeq {m_ec^2 \over \gamma c \sigma_T u_{ext}'} \, 
\ee
with the dynamical time scale of the emission region $t_{dyn}'$, one
finds the critical electron Lorentz factor
\be 
\gamma_c \simeq {m_e c^2 \Gamma \over \sigma_T u_{ext}' r} \, , 
\ee
where we assumed that the energy density of the external photon field may be
approximated as being constant up to some characteristic distance $r_{ext}$,
and that the blazar emission zone is located at $r_0 < r \le r_{ext}$, where
$r_0$ is the distance scale of the jet acceleration (see \S\,{3.1.2}). 

In the case of the dominant EC(BLR) emission, i.e. when the electron
cooling is dominated by Comptonization of the photon field due to the
broad emission lines, we have $r_0 < r \le r_{ext}=r_{BLR}$, where
\be 
r_{BLR} \simeq 0.1 \, \left({L_{disk} \over 10^{46}\,{\rm erg\,s^{-1}}}\right)^{1/2} 
\,\, {\rm pc} \, 
\ee
(Ghisellini \& Tavecchio 2008 and references therein), and
\be
\label{u_blr}
u_{ext}' \simeq u_{BLR} \Gamma^2 \simeq 
{\xi_{BLR} L_{disk} \Gamma^2 \over 4 \pi r_{BLR}^2 c} \simeq 
12 \, \left({\xi_{BLR} \over 0.1}\right) \, \left(\Gamma \over 20\right)^2 
\,\,{\rm erg\,cm^{-3}} \, .
\ee
If, instead, $r_{BLR} < r < r_{HDR}$, where
\be
\label{r_hdr}
r_{HDR} \simeq 4 \, \left({L_{disk} \over 10^{46}\,{\rm
    erg\,s^{-1}}}\right)^{1/2} \left({T \over 10^3\,{\rm
    K}}\right)^{-2.6} \,\, {\rm pc} \, ,
\ee
the electron cooling is dominated by IC up-scattering of the 
near-IR photons emitted by the hot dust with the temperature $T$, located 
in the molecular torus and irradiated by the accretion disk (Nenkova et al. 
2008 and references therein).  In such a case
\be
\label{u_hdr}
u_{ext}' \simeq u_{HDR} \Gamma^2 \simeq {\xi_{HDR} L_{disk} \Gamma^2 \over 4 \pi
  r_{HDR}^2 c} \simeq 8 \times 10^{-3} \, \left({T \over 10^3\,{\rm
    K}}\right)^{5.2} \left({\xi_{HDR} \over 0.1}\right) 
\, \left(\Gamma \over 20\right)^2 \,\,{\rm erg\,cm^{-3}} \, ,
\ee
and the EC(HDR) emission dominates the high-energy radiative output of
the source.

With all the values quoted above, the critical electron Lorentz factor
reads as
\be 
\gamma_c \simeq \left({\Gamma \over 20}\right)^{-1} \left({L_{disk} \over 
10^{46}\,{\rm erg\,s^{-1}}}\right)^{-1/2}  \times \left\{ \begin{array}{cc}
7 \, \left({r / r_{BLR}}\right)^{-1} \left({\xi_{BLR} / 0.1}\right)^{-1} & \\
250 \, \left({r / r_{HDR}}\right)^{-1} \left({\xi_{HDR} / 0.1}\right)^{-1} 
\left({T / 10^3\,{\rm K}}\right)^{-2.6} & 
\end{array} \right. \, , 
\ee
and the corresponding break frequency in the high-energy spectral
component, $\nu_{EC,\,c} \simeq \gamma_c^2 \Gamma^2 \nu_{ext}$, may be
evaluated as
\be 
\left({h\nu_{EC,\,c} \over {\rm MeV}}\right) \simeq \left({L_{disk} \over 
10^{46}\,{\rm erg\,s^{-1}}}\right)^{-1} \times \left\{ \begin{array}{cc} 
0.2 \, \left({r / r_{BLR}}\right)^{-2} \left({\xi_{BLR} / 0.1}\right)^{-2} & \\
8 \, \left({r / r_{HDR}}\right)^{-2} \left({\xi_{HDR} / 0.1}\right)^{-2}  
\left({T / 10^3\,{\rm K}}\right)^{-5.2} 
&  \end{array} \right. \, ,
\ee
where $h \nu_{ext}$ has been substituted with $h\nu_{BLR}\simeq
10$\,eV and $h\nu_{HDR} \simeq 0.3$\,eV, respectively.  Hence, the
production of high-energy emission in the EC(HDR) model proceeds in
the slow cooling regime up to the MeV photon energy range, in
agreement with the position of luminosity peaks observed in luminous 
blazars.  In the EC(BLR) model the $\gamma$-ray spectra are expected to
soften below the MeV range, but still at sufficiently high photon energies 
to assure consistency with observations.
Hence, in similarity to the case of the SSC model, the observed 
very hard X-ray spectra of powerful blazar sources are in agreement 
with the EC scenario for the high-energy jet emission.  

High energy radiation produced in the innermost portions of AGN jets may also 
be contributed by Comptonization of the direct emission of the accretion 
disk illuminating the blazar emission zone from the sides and from behind. 
This contribution should in particular dominate the high-energy radiative output
of the jet at distances
\be 
r < r_{disk/BLR} \simeq 0.6 \, \left({r_{BLR}^2 \, R_g \over \xi_{BLR}}\right)^{1/3} 
\simeq 0.01 \, \left(\xi_{BLR} \over 0.1 \right)^{-1/3}
\left(L_{disk} \over 10^{46}\,{\rm erg \, s}^{-1} \right)^{1/3}
\left( M_{BH} \over 10^9 M_{\odot}\right)^{1/3} \, {\rm pc} \, 
\ee 
(Dermer \& Schlickeiser 2002). Energy dissipation events taking place on 
such small scales may be responsible for production of intraday flares observed
sporadically in some blazar sources (see \S\, {7.2}).  

\section{Synchrotron and Inverse-Compton Peak Frequencies}

In the case of luminous blazars, the observed synchrotron peak
frequencies seem to cluster within a relatively narrow frequency range
$10^{12}-10^{14}$\,Hz (Ghisellini \& Tavecchio 2008 and references
therein). Meanwhile, the peak of the high-energy spectral components
of the objects discussed here have to be located between $0.1$\,MeV
and $0.1$\,GeV photon energies, as indicated by the observed values
of the X-ray and $\gamma$-ray spectral indices (e.g., Fossati et
al. 1998, Ghisellini \& Tavecchio 2008).  Again, the relative location
of the low- and high-energy spectral peaks does not provide any
constraints on the hadronic models, because the poorly known particle
acceleration efficiency (the $f$ parameter introduced in Eq.~\ref{t_acc}
above) is claimed to be substantially different for electrons
and protons (e.g., Reimer et al. 2004), and therefore the ratio of
peak frequencies is in fact a free parameter.
However, in the leptonic scenarios for the
high-energy blazar emission, the relative location of spectral peaks
offers interesting constraints on the model parameters, as discussed
below.

\subsection{Peak Frequencies in the SSC Model}

Peak frequencies of the synchrotron and SSC components in the spectra
of blazar sources are determined by the critical (break) energies of
electrons emitting bulk of the emission,
\be 
\gamma_{br} \simeq \sqrt{\nu_{SSC,\,br} \over \nu_{syn,\,br}} \, . 
\ee
When combined with the standard formulae for the corresponding break
frequency of the synchrotron photons
\be 
\nu_{syn,\,br} \simeq 4 \times 10^6 \, B_{\rm G}' \gamma_{br}^2 
\Gamma \,\, {\rm Hz} \, , 
\ee
we obtain
\be h \nu_{SSC, \, br} = 10^{18} \, {B_{\rm G}'}^{-1}
\left({\nu_{syn,\,br} \over 10^{13}\,{\rm Hz}}\right)^2 \left({\Gamma
  \over 20}\right)^{-1} \,\, {\rm Hz} \, .  
\ee
In the SSC model, magnetic field intensity $B'$ given by the relation
$u_{syn}'/u_B' \simeq q$ is
\be B_{\rm G}' \simeq 10^{-2} \, \left({L_{syn} \over 10^{47}\,{\rm
    erg\,s^{-1}}}\right)^{1/2} \left({\Gamma \over 20}\right)^{-1}
\left({r \over {\rm pc}}\right)^{-1} \left({q \over 10}\right)^{-1/2}
\, , \ee
and hence
\be h\nu_{SSC, \, br} \simeq 0.4 \, \left({\nu_{syn,\,br} \over
  10^{13}\,{\rm Hz}}\right)^2 \left({\Gamma \over 20}\right) \,
\left({r \over {\rm pc}}\right) \, \left({q \over 10}\right)^{1/2}
\left({L_{syn} \over 10^{47}\,{\rm erg\,s^{-1}}}\right)^{-1/2}
\,\,{\rm MeV} .  \ee
This would then be consistent with the observational constraints for
the location of the emitting zone $r>$ a few pc.

However, the SSC mechanism can dominate production of the $\gamma$-ray
emission only if the energy density of synchrotron radiation is larger
than energy density of the external radiation field, both as measured
in the source co-moving frame, i.e. if $u_{syn}' > u_{ext}' \simeq
u_{ext} \Gamma^2$.  This condition can be satisfied on parsec scales,
provided the contribution from the BLR at these distances is already
negligible (as expected even in luminous blazars) and the dust
temperature is much below the sublimation temperature.  The second
requirement can be verified observationally by investigating the
presence of a hot dust in mid- and near-IR data for non-blazar
radio-loud quasars, i.e.\ for jet sources observed at $\theta_{obs}
\gg 1/\Gamma$.  According to Cleary et al. (2007), who analyzed
Spitzer observations of 3C quasars, the dust temperatures are in the
range $T\sim 100-300$\,K.  Such temperatures are derived assuming that
the observed mid-IR excess in the spectra of analyzed objects is
produced by a high energy tail of the synchrotron radiation of a large
scale jet.  If this is indeed the case, then there is a distance range
between $r_{BLR}$ and $r_{HDR}$ where the SSC process can dominate the
production of $\gamma$-rays over the EC(HDR) mechanism.  On the other
hand, we note that the near-IR quasar continua join the harder optical
ones almost always around the same frequency, namely $\sim 1$\,$\mu$m,
in \emph{both} radio-quiet and radio-loud sources (Barvainis 1987;
Glikman et al. 2006; Netzer et al. 2007; Labita, Treves, \& Falomo
2008).  Therefore, the assumption regarding the synchrotron origin of
near IR radiation seems not to be justified, requiring in addition
very precise tuning of the jet parameters.  At the same time, a
natural explanation for the observed spectral dip around $1$\,$\mu$m
frequencies in quasar spectra is provided by the clumpy torus model,
in a framework of which the near-IR emission is produced by a hot dust
with the temperature close to the sublimation value, $T \sim 1500$\,K
(Nenkova et al. 2008).
 
\subsection{Peak Frequencies in the EC Models} 

In the EC models, the high energy (inverse-Compton) spectral component 
is expected to peak at the observed frequencies
\be 
\nu_{EC,\,br} \simeq \gamma_{br}^2 \Gamma^2 \nu_{ext} \, , 
\ee
and the ratio of the peak frequencies reads as
\be {\nu_{EC, \, br} \over \nu_{syn,\, br}} \simeq 10^{9} \,
\left({\Gamma \over 20}\right) \, \left({h\nu_{ext} \over {\rm
    eV}}\right) \, {B_{\rm G}'}^{-1} \, .  
\ee
With the magnetic field given by the $q=\Gamma^2 u_{ext}/u_B'$
condition,
\be
B_{\rm G}' = \Gamma \, \sqrt{{8\pi \over q}} \, \left({u_{ext} 
\over {\rm erg\,cm^{-3}}}\right)^{1/2} \simeq 
\left({\Gamma \over 20}\right) \, \left({q \over 10}\right)^{-1/2} \times 
\left\{ \begin{array}{cc} 5 \, \left({\xi_{BLR} / 0.1}\right)^{1/2} & \\
0.1 \, \left({\xi_{HDR} / 0.1}\right)^{1/2} \left({T / 10^3\,{\rm K}}\right)^{2.6} 
 & \end{array} \right. \quad , 
\ee
this ratio is then surprisingly similar for both cases considered
here, EC(BLR) and EC(HDR):
\be
\label{nu_br_ratio}
{\nu_{EC,\,br} \over \nu_{syn,\,br}} \simeq 3 \times 10^9 \,
\left({q \over 10}\right)^{1/2} \times \left\{ \begin{array}{cc}
  \left({\xi_{BLR} / 0.1}\right)^{-1/2} & \\ \left({\xi_{HDR} /
    0.1}\right)^{-1/2} \left({T / 10^3\,{\rm K}}\right)^{-2.6} &
\end{array} \right. \, , 
\ee
leading to the expected spectral location of the EC luminosity peak
\be h\nu_{EC,\, br} \simeq 0.1 \, \left({q \over 10}\right)^{1/2}
\left({\xi_{ext} \over 0.1}\right)^{-1/2} \left({\nu_{syn,\,br} \over
  10^{13}\,{\rm Hz}}\right) \,\, {\rm GeV} \, , 
\ee
with the assumed $T \sim 10^3$\,K.  The $h\nu_{EC,\, br}$ evaluated 
above seems to be somewhat higher than the peak photon energies of the
high-energy continua claimed for luminous blazars (Ghisellini \&
Tavecchio 2008), as long as $\nu_{syn,\,br}$ is not smaller than
$10^{13}$\,Hz. However, we note that the high energy peak frequencies
are not observed directly, but only reconstructed from the power-law
slopes of the lower-energy (X-ray) and higher-energy (GeV) segments of
the IC continua, assuming a single-broken power-law
spectral shape of the emission.  Meanwhile, if the injected electron
spectrum is flatter than $\gamma^{-2}$ below electron energies
$\gamma_{br}$ corresponding to the peak frequencies $\nu_{syn,\,br}$
and $\nu_{EC,\,br}$ (as seems to be indicated by the observed hard
X-ray spectra with $\alpha_x < 0.5$), and steeper than that above
$\gamma_{br}$ (see in this context Ghisellini \& Tavecchio 2008, but
also \S\,9 of this paper), the `reconstructed' peak frequency can
easily be lower than the `true' break frequency $\nu_{EC,\,br}$ by a
factor of $(\nu_{EC,\,c} / \nu_{EC,\,br})^{(0.5 - \alpha_x) /
  (\alpha_{\gamma} - \alpha_x)}$, i.e. by possibly as much as $\sim
10$ for the typical values of the X-ray and GeV spectral indices of
luminous blazars (see Table~1).

\section{Lack of Bulk-Compton Spectral Features}

In astrophysical jets, \emph{all} ultrarelativistic particles cool due
to radiative energy losses and also due to adiabatic expansion of the
emitting regions.  Hence, the observed high-energy non-thermal
radiation requires \emph{in situ} particle acceleration and one may
expect to observe at any given instant radiation from both 'cold' and
'hot' population of particles.
Typically, jet electrons constituting a cold population are considered
to be non-relativistic in the jet rest frame in the innermost parts of
the outflow (say, hundreds of Schwarzschild radii from the active
center), because of the extremely rapid (catastrophic) particle
cooling at the jet base. Such electrons, if indeed present in a steady
flow, are predicted to produce distinct bulk-Compton features in
blazar spectra, most pronounced around $h\nu_{bc} \sim \Gamma^2
h\nu_{ext}$ photon energies, i.e. around $\sim 1-4$\,keV for $\Gamma
\sim 10-20$ and $h\nu_{ext} \sim 10$eV (Begelman \& Sikora 1987;
Sikora et al. 1997; Sikora \& Madejski 2000). In addition, in the
internal shock model (Spada et al. 2001), presence of cold jet
electrons should manifest itself as soft and hard X-ray precursors of
non-thermal blazar flares (Moderski et al. 2004). Both those `steady'
and 'dynamic' bulk-Compton features are expected to be particularly
strong if produced close to the black hole, in the dense radiation
field of an accretion disk. So far, no clear detection of such features 
has been reported, with the possible exception of weak soft X-ray excesses 
claimed in the spectra of some blazars (see, e.g., Worsley et al. 2004; 
Kataoka et al. 2008; De Rosa et al. 2008). Such an apparent weakness of 
bulk-Compton features may result from the fact that in the very central 
regions of AGN jet the (magnetohydrodynamical) acceleration of the outflow 
is not yet fully complete (as noted previously in Sikora et al. 2005; Celotti, 
Ghisellini \& Fabian 2007). On the other hand,
analogous spectral features produced further out at the point where
the jet is already accelerated to terminal bulk velocities are predicted
to be still detectable, in particular if the blazar emission zone is
enclosed within the BLR.

However, it should be emphasized that the involved assumptions
regarding cold electron population may not be valid at all
scales. Namely, at some larger distances from the nucleus, the
radiative and adiabatic cooling are expected to become inefficient
when compared to the acceleration processes at very low electron
energies. As a result, shock heating and/or interaction of particles
with magnetic turbulence (driven inevitably by MHD instabilities of a
magnetized outflow, and the related magnetic reconnection processes)
may effectively accelerate all the cold electrons to energies $\gamma
\gg 1$, replacing the non-relativistic Maxwellian-like electron
population with a very hard, quasi-power-law distribution that joins
smoothly the higher-energy electron tail (see \S\,9 below).  In such a
case, bulk-Compton features are expected to be absent.  Yet another
alternative explanation for the apparent lack of radiative signatures
of cold electrons in blazar spectra is that formation and acceleration
of jets in luminous sources is still not completed within the BLR.  If
so, bulk-Compton features produced at $r < r_{BLR}$ may be simply too
weak to be distinguished from the broad-band continuum generated via
EC(HDR) process. The latter possibility is supported by the fact that
in several cases pure power-law X-ray continua extend down to energies 
$0.1-0.3$\,keV, with no evidence for any soft excess (e.g. in 3C\,279: 
see Lawson \& M$^c$Hardy 1998; Collmar et al. 2007).

\section{Lack of Klein-Nishina Spectral Features}

Recent studies of the Klein-Nishina (KN) effects of relativistic sources
immersed in a dense external radiation field and dominated by the IC
emission show that for the maximum electron energies $\gamma_{max} \gg
\gamma_{KN} \equiv m_e c^2 / 4\Gamma h\nu_{ext}$ the IC continuum
should break at much larger frequencies than $\nu_{KN} = \gamma_{KN}^2
\Gamma^2 \nu_{ext}$, and that the synchrotron component should harden
around
\be
\nu_{syn,\,KN} \simeq 4 \times 10^6 \, \gamma_{KN}^2 \Gamma \, B_{\rm G}' \,{\rm Hz}
\simeq 3 \times 10^{15} \, B_{\rm G}' \, \left({\Gamma \over 20}\right)^{-1}  
\left({h \nu_{ext} \over {\rm eV}}\right)^{-2} \,\, {\rm Hz} \, . 
\ee
(Dermer \& Atoyan 2002; Moderski et al. 2005). Both these effects
result from hardening of the injected electron energy distribution due
to the reduction of the IC cooling rate in the KN regime,
i.e. for the electrons with $\gamma > \gamma_{KN}$. In a framework of
the EC(BLR) model, the related spectral features are expected to be
pronounced already for electrons with energies $\gamma > 3 \times
10^3$, and the hardening of the synchrotron continuum (being
especially strong for mono-energetic external radiation fields) is
predicted to be visible already in the optical band. However, as shown
by Tavecchio \& Ghisellini (2008), emission of BLR is not exactly
monochromatic, and in particular is characterized by a significant
low-frequency power-law tail. As a result, the expected KN hardening
in the optical band may be relatively weak, though possibly still
prominent in the UV band. The dependence of such KN effects on the
maximum electron energy and the parameter $q$ for the EC(BLR) model is
illustrated in Figures~1 and~2, with an assumed electron energy distribution
$N_{\gamma} \propto \gamma^{-s}$ with $s = 2.4$. As shown in these
Figures, high-$q$ blazars with synchrotron and IC spectra extending
respectively above $1$\,eV and $10$\,GeV photon energies should
display in their broad-band continua the KN spectral features
discussed above, \emph{unless} the production of $\gamma$-rays is dominated
by Comptonization of IR radiation of hot dust, 
and \emph{only if} the electron energy distribution
extends to  $\gamma > 3 \times 10^3$ with the power-law slope not much
steeper than $s \simeq 2.5$. 
This provides a possibility to determine
the  characteristic energies of the dominant seed photons via 
analysis of the Klein-Nishina effects, 
manifested by different spectral shapes of the high energy tails of 
the synchrotron and EC continua.  

\section{EC(HDR) versus EC(BLR) Models}

As shown in \S\,4, the same luminosity ratio of the IC and synchrotron
spectral components implies similar high-energy peak frequencies in
both EC(BLR) and EC(HDR) models. And this is clearly coincidental
since the location of the two spectral peaks happens to scale with a
factor $\nu_{ext}/\sqrt{u_{ext}}$ (see \S\,3.3).  Interestingly, the
conspiracy between the two models is reflected also by several other --
though not by all -- important model output parameters, as discussed
below.  

\subsection{Similarities}

Let us consider a `fiducial' source located within a given distance
range $\Delta r \sim r$, and immersed in a given external radiation
field. The observed X-ray and $\gamma$-ray spectral indices of such a source,
$\alpha_x$ and $\alpha_{\gamma}$ respectively, allow one to
reconstruct the energy distribution of radiating electrons,
approximated here for simplicity by a single-broken power-law. In
particular, using the approximate formula
\be L_{\nu}d\nu \simeq N_{\gamma} d\gamma \, |\dot \gamma| \, {\rm m_e
  c^2} \, \Gamma^4 \, 
\ee
we find that 
\be 
N_{\gamma} = C_{N_1}\gamma^{-s_1}(1+\gamma/\gamma_{br})^{s_1-s_2} \, , 
\ee
\be 
\gamma_{br}= \left(C_{N_2} \over C_{N_1}\right)^{1/(s_1-s_2)} \, , 
\ee 
\be 
C_{N_1} \simeq {2 \nu_{ext}^{1-\alpha_x} \over 
c \sigma_T u_{ext} \Gamma^{4+2\alpha_x}} \, C_x \, , 
\ee
\be 
C_{N_2} \simeq {2 \nu_{ext}^{1-\alpha_{\gamma}} \over 
c \sigma_T u_{ext} \Gamma^{4+2\alpha_{\gamma}}} \, C_{\gamma} \, , 
\ee
\be 
C_x = {[\nu_x L_{\nu_x}] \over \nu_x^{1-\alpha_x}}  \,
= {4 \pi d_L^2 [\nu_{obs,\,x} F_{\nu_{obs,\,x}}] \over \nu_{obs,\,x}^{1-\alpha_x}} \, , 
\ee
\be 
C_{\gamma} = {[\nu_{\gamma} L_{\nu_{\gamma}}] \over 
\nu_{\gamma}^{1-\alpha_{\gamma}}}
= {4 \pi d_L^2 [\nu_{obs,\,\gamma} F_{\nu_{obs,\,\gamma}}] \over 
\nu_{obs,\,\gamma}^{1-\alpha_{\gamma}}} \, , 
\ee
\be 
s_1= 2\alpha_x +1 \,\, , \, {\rm and} \,\, s_2=2\alpha_{\gamma} +1 \, . 
\ee
Here $F_{\nu_{obs}}$ are the observed monochromatic energy fluxes at
some particular observed frequency $\nu_{obs}$, $d_L$ is the
luminosity distance of the source, and $z$ is its redshift.
 
The number of electrons contributing to the radiation at a given
instant of observation can be approximated by
\be N_e=\int_1{N_{\gamma} \, d\gamma}\simeq {C_{N_1} \over 2 \alpha_x}
\, , 
\ee
as long as $\gamma_{br} \gg 1$ and $\gamma_{min}=1$.
Hence, the electron number flux is
\be \dot N_e \simeq {N_e \over (R/\Gamma c)} \simeq {\Gamma^2 N_e
  \over (r_{ext}/c)} \propto {\nu_{ext}^{1-\alpha} \over u_{ext} \,
  r_{ext}}\, , 
\ee
and the ratio of electron energy fluxes in the two considered models
reads as
\be 
{\dot N_{e,\,HDR} \over \dot N_{e, \, BLR} } \simeq  
\left(\nu_{HDR} \over \nu_{BLR}\right)^{1-\alpha_x}
\left(u_{HDR} \over u_{BLR}\right)^{-1} \left(r_{HDR} \over r_{BLR}\right) 
\simeq 6 \, \left({T \over 10^3\,{\rm K}}\right)^{-2.6} \left(\xi_{HDR} 
\over \xi_{BLR}\right)^{-1} \, ,
\ee
which, for the sublimation temperature $T \simeq 1500$K and $\xi_{HDR}
\sim 3 \, \xi_{BLR}$, is of the order of unity.

Assuming further that the energy flux of a jet is dominated by
nonrelativistic protons, the proton number flux is simply
\be 
\dot N_p \simeq {L_{j} \over  \Gamma  m_p c^2 } \, , 
\ee
and therefore the pair content in the two models is
\be 
{(n_e / n_p)_{HDR} \over (n_e/n_p)_{BLR} } = 
{\dot N_{e,\,HDR} \over \dot N_{e,\,BLR}} \, , 
\ee
i.e. almost exactly the same. In addition, similar value for
$\dot N_e$ derived for the two models implies also similar electron
energy fluxes,
\be 
L_e = \langle \gamma \rangle \, m_ec^2 \, \Gamma \dot N_e \, , 
\ee
where $\langle \gamma \rangle = \int{N_{\gamma}\gamma\, d\gamma} /
\int{N_{\gamma}}\, d\gamma \simeq \ln{\gamma_{br}}$ is the average
electron Lorentz factor (assuming $\alpha_x \simeq 0.5$), and
$\gamma_{br} = \Gamma^{-1} \sqrt{\nu_{EC,\,br}/\nu_{ext}}$. It follows
directly from the above discussion (\S\,3 and \S\,4) that $\langle
\gamma \rangle$, and thus $L_e$, is the same for the EC(BLR) and
EC(HDR) models. Finally, also magnetic energy fluxes $L_B \simeq c
u_B' \Gamma^2 \pi R^2 \sim c u_B' \pi r^2$ do not differ in both
cases considered above, namely
\be 
{L_{B,\,HDR} \over L_{B,\,BLR}} = 
{u_{HDR} \, r_{HDR}^2 \over u_{BLR}\, r_{BLR}^2} = 
0.65 \, \left({T \over 10^3\,{\rm K}}\right)^{2.6} {\xi_{HDR} \over \xi_{BLR}} \, . 
\ee

\subsection{Differences}

Certainly there are some aspects which differentiate the two models
considered.  These are (i) the critical synchrotron self-absorption
frequencies, (ii) the characteristic variability time scales, and
(iii) the low energy breaks/cut-offs of the IC spectral component.

(i) The synchrotron self-absorption makes the source optically thick
below the frequency
\be 
\nu_a \simeq 4 \times 10^{11} \, {B_{\rm G}'}^{1/7} \left({[\nu_a L_{\nu_a}] \over 
10^{47}\,{\rm erg\,s^{-1}}}\right)^{2/7} 
\left({\Gamma \over 20}\right)^{3/7} \left({r \over {\rm pc}}\right)^{-4/7}
\,\, {\rm Hz} \, , 
\ee
which may be re-written as
\be 
\left({\nu_a \over 10^{12}\,{\rm Hz}}\right) \simeq 
\left({[\nu_a L_{\nu_a}] \over 10^{47}\,{\rm erg\,s^{-1}}}\right)^{2/7} \times
\left\{ \begin{array}{cc}
2 \, \left({B' / 5\,{\rm G}}\right)^{1/7} \left({r / r_{BLR}}\right)^{-4/7} & \\
0.2 \, \left({B' / 0.1\,{\rm G}}\right)^{1/7} \left({r / r_{HDR}}\right)^{-4/7} & 
\end{array}\right. \, . 
\ee
Meanwhile, millimeter and submillimeter observations indicate
absorption break in blazar spectra around $1$\,mm (see, e.g., Gear et
al. 1994). This favors the EC(HDR) model over the EC(BLR) version.

(ii) The shortest time scales of flares differ in the two EC 
models by a factor
$r_{HDR}/r_{BLR} \sim 30 \, (T/10^3\,{\rm K})^{2.6}$. In particular, one has
\be \left({t_{min} \over {\rm day}}\right) \simeq \left({L_{disk} \over
  10^{46}\,{\rm erg\,s^{-1}}}\right) \, \left({\Gamma \over
  20}\right)^{-2} \times \left\{ \begin{array}{cc} 0.4 \, & \\ 10 \,
  \left({T / 10^3\,{\rm K}}\right)^{-2.6} &
\end{array} \right. \, . 
\ee 
Hence, the characteristic variability time scale predicted by the
EC(HDR) model is more consistent with the one implied by the observed
variability patters in `optically violent variable' quasars, which is
of the order of days/weeks  (Pica et al. 1988)  
A nice example where such time scales are seen both
in the $\gamma$-ray and IR/optical frequency ranges is provided by
multiwavelength observations of blazar 3C~454.3 (see Fig.~1 in
Bonning et al. 2009).

We note in this context that in some cases rapid, day/intraday flares
are imposed on the observed light-curves of blazar sources.  One
example is the extremely rapid ($\sim 2$ hours) optical flare, as
recorded also in 3C~454.3 (Raiteri et al. 2008), or intraday
$\gamma$-ray flare recorded by {\it Fermi}/LAT in the radio quasar PKS
1454-354 (Abdo et al. 2009a). Such rapid flares may indicate that in
addition to the main `energy dissipation site' located at parsec-scale
distances from the active center, there are also episodic dissipation
events taking place on sub-parsec distances, possibly related to MHD
instabilities and/or internal shocks within the inner parts of a
strongly magnetized outflow.
 
(iii) Because the  spectra produced by relativistic electrons emerge
at energies $h\nu_{low} \simeq \Gamma^2 \nu_{ext}$, in the case of the
EC(BLR) model a low energy break/cut-off should be observed around
photon energies $\sim 4\,(\Gamma/20)^2$\,keV, while in the case of the
EC(HDR) should be at around $\sim 0.1\,(\Gamma/20)^2$\,eV (i.e., basically
below the low-energy threshold of the available X-ray instruments).
Noting however that the X-rays can be contributed by several other
processes (such as SSC emission, bulk-Compton radiation, or
high-energy emission of accretion disk corona), the apparent lack of
the expected break can be used as an argument in favor of the EC(HDR)
model only with caution.

\section{Hadrons in the EC Models}

If the EC models are correct, then the broad-band blazar spectra allow
us to estimate the number of jet electrons down to lowest energies,
and to find the corresponding particle flux. However, since (by
definition) in the `leptonic models' the entire observed radiative
output of a jet is produced by primary (i.e., directly accelerated)
electrons, the proton content of an outflow remains in principle
unconstrained. The most recent analysis indicate that it cannot be
negligible, though. In fact, several different authors came to the
conclusion that the cold protons indeed dominate the jet bulk kinetic
power, at least at $> r_{BLR}$ scales, and that the number ratio of
electron-proton to electron-positron pairs in quasar jets is of the
order of $0.1-1$ (Sikora \& Madejski 2000, Celotti \& Ghisellini
2008).  Uncertainties regarding the proton content could be solved
directly provided acceleration of protons up to ultra-relativistic
energies indeed takes place in the innermost parts of quasar jets with
high efficiency, and is followed by production of some distinctive
electromagnetic features, such as hardening of $\gamma$-ray spectra
above $10$\,GeV or softening of X-ray spectra at lower
energies. Again, lack of such spectral features does not exclude the
dynamical role of protons in general, but only limits the number
of ultrarelativistic hadrons, and thus constrains the particle
acceleration models rather than the jet content.

In order to discuss this issue in more detail, we estimate the
efficiency of the photo-meson production process for distances and
magnetic fields appropriate for the EC models of luminous
blazars. With the target photons provided by the local synchrotron
emission of directly accelerated electrons, we have the appropriate
opacity
\be \tau_{p\gamma}^{(syn)}(\gamma_p) \simeq \langle\sigma_{p\gamma}
K_{p\gamma}\rangle \, {L_{syn} \gamma_p \, \zeta\!(\gamma_p) \over 4
  \pi r \, \Gamma^3 \, m_{\pi} {\rm c^3}} \, . \ee
where $\zeta(\gamma_p) < 1$ depends on the shape of the synchrotron
spectrum (see \S{3.1}).  Assuming further that protons are injected with 
the same
power-law index as directly accelerated jet electrons producing the
observed $\gamma$-ray emission in the fast cooling regime, we have
$s_p = 2 \alpha_{\gamma} \ge 2$ for the typical $\gamma$-ray spectra
of luminous blazars $\alpha_{\gamma} \ge 1$ (Abdo et al. 2009b).
Hence, because in addition $\eta_{p\gamma}^{(syn)}|_{s_p >2} >
\eta_{p\gamma}^{(syn)}|_{s_p =2}$ and $\zeta(\gamma_p)<1$, we find
using Eq.~(\ref{eta_pgamma}) that
\be 
\eta_{p\gamma}^{(syn)} < \langle \sigma_{p\gamma} K_{p\gamma}\rangle \, 
{L_{syn}  \over 4 \pi r \, \Gamma^3 \, m_{\pi} {\rm c^3}} \, 
{\gamma_{p,\,max} \over \ln{\gamma_{p,\,max}}} \, 
\ee
which, for
\be \gamma_{p,\,max} \simeq 5 \times 10^{9}\, B_{\rm G}' \, \left({f
  \over 10}\right)^{-1} \left(r \over {\rm pc}\right) \, \left(\Gamma
\over 20\right)^{-1} \simeq 3 \times 10^{9} \, \left({f \over
  10}\right)^{-1} \left(L_B \over 10^{46}\,{\rm erg\,s^{-1}} \right)
\, \left( \Gamma \over 20 \right)^{-1} 
\ee
(see \S\,3.1), gives finally
\be 
\eta_{p\gamma}^{(syn)} < 4 \times 10^{-4} \left({f \over 10}\right)^{-1}\!
\left(L_{syn} \over 10^{47}\,{\rm erg \, s^{-1}} \right)\!
\left(L_B \over 10^{46}\, {\rm erg \, s^{-1}} \right)^{1/2}\!   
\left(\Gamma \over 20 \right)^{-4}\! \left(r \over {\rm pc}\right)^{-1}\! 
\left(\ln{\gamma_{p,\,max}} \over 24\right)^{-1} . 
\ee
The particular value of the magnetic energy flux $L_B \simeq
10^{46}$\,erg\,s$^{-1}$ anticipated above, is justified by
Eqs.~(\ref{u_blr}), (\ref{u_hdr}), and (\ref{nu_br_ratio}), which in turn give
\be {L_B \over L_{disk}} \simeq \left({\Gamma \over 20}\right)^2 \,
\left({q \over 10}\right)^{-1} \times \left\{ \begin{array}{cc} 0.9 \,
  \left({\xi_{BLR} / 0.1}\right) & \\ 0.5 \, \left({\xi_{HDR} /
    0.1}\right) \, \left({T / 10^3\,{\rm K}}\right)^{2.6}
  & \end{array} \right. \quad .  \ee
As shown above, in the EC models for luminous blazars, high efficiency of
the photo-meson production is expected only at distances $r \ll 1 {\rm pc}$. 
This implies that the maximal hadronic radiative output is
reachable at around $r \sim r_0$, where the jet is already fully
developed (i.e., accelerated and collimated), and the Doppler boosting
is large. 
This efficiency can approach 
$\sim  \, (f/10)^{-1}(r_0/0.03\,{\rm pc})^{-1}\%$. 
Hence, even in this most optimistic case, the efficiency of the 
photo-meson production is much lower than the 
$\sim 50\%$ radiative efficiency of jet electrons.  

Let us further evaluate the number densities of the external (BLR
and HDR) photon fields in the jet rest frame, approximating these by
a monoenergetic distributions with the characteristic comoving photon
energies $h\nu_{BLR}' \simeq 200 (\Gamma/20)$\,eV and $h\nu_{HDR}'
\simeq 6(\Gamma/20)$\,eV, respectively:
\be 
n_{BLR}' = {u_{BLR}' \over h \nu_{BLR}'} \simeq
3 \times 10^{10} \, \left(\xi_{BLR} \over 0.1\right)\,
\left(\Gamma \over 20 \right) 
\ee
at $r_0 < r < r_{BLR}$, and
\be n_{HDR}' = {u_{HDR}' \over h \nu_{HDR}'} \simeq
10^9 \, \left(T \over 10^3\,{\rm K}\right)^{5.2} \, 
\left(\xi_{HDR} \over 0.1 \right) \, \left(\Gamma \over 20 \right) 
\ee
at $r_{BLR} < r < r_{HDR}$. For $\gamma_p > \gamma_{p,\,th}$, where
\be 
\gamma_{p,\,th}^{(BLR)} \simeq {m_{\pi}c^2 \over h\nu_{BLR}'} \simeq 
7 \times 10^5 \, \left(\Gamma \over 20\right)^{-1} \, ,
\ee
and
\be 
\gamma_{p,\,th}^{(HDR)} \simeq {m_{\pi}c^2 \over h \nu_{HDR}'} \simeq
2 \times 10^7 \, \left(\Gamma \over 20 \right)^{-1} \, ,
\ee
we have 
\be 
\tau_{p\gamma}^{(BLR)} \simeq \langle\sigma_{p\gamma}K_{p\gamma}\rangle \,
{n_{BLR}' \, r \over \Gamma} \simeq
4 \times 10^{-2} \, \left(\xi_{BLR} \over 0.1\right) \,
\left(r \over r_{BLR}\right)\, 
\left(L_{disk} \over 10^{46}\,{\rm erg \, s^{-1}}\right)^{1/2} \, , 
\ee
and
\be 
\tau_{p\gamma}^{(HDR)} \simeq \langle\sigma_{p\gamma}K_{p\gamma}\rangle \,
{n_{HDR}' \, r \over \Gamma} \simeq
4 \times 10^{-2} \, \left(\xi_{HDR} \over 0.1\right) \,
\left(r \over r_{HDR}\right) \, 
\left(L_{disk} \over 10^{46} \,{\rm erg \, s^{-1}}\right)^{1/2} 
\left(T \over 10^3 \,{\rm K}\right)^{2.6} \, .
\ee
Hence, this implies the total photo-meson production efficiency to be 
\be 
\eta_{p\gamma}^{(BLR)} \le 4 \times 10^{-2} \,
\left(\xi_{BLR} \over 0.1\right) \, \left(r \over r_{BLR}\right) \,
\left( L_{disk} \over 10^{46} \, {\rm erg \, s^{-1}} \right)^{1/2}
{\ln{(\gamma_{p,\,max}/\gamma_{p,\,th}^{(BLR)})} \over \ln{\gamma_{p,\,max}}} 
\ee
for $r_0 < r < r_{BLR}$, and
\be 
\eta_{p\gamma}^{(HDR)} \le 4 \times 10^{-2} \,
\left(\xi_{HDR} \over 0.1\right) \, \left(r \over r_{HDR}\right) \,  
\left( L_{disk} \over 10^{46}\,{\rm erg \, s^{-1}} \right)^{1/2}
\left(T \over 10^3\,{\rm K}\right)^{2.6} \,
{\ln{(\gamma_{p,\,max}/\gamma_{p,\,th}^{(HDR)})} \over \ln{\gamma_{p,\,max}}}
\ee
for $r_{BLR} < r < r_{HDR}$.  Note that the calculated efficiencies
never exceed $\sim 1\%$.

Summarizing this section, we conclude that some imprints of hadronic
  activity in the EC-dominated high-energy blazar spectra are
  anticipated only if the blazar zone is located at $r_0 < r <
  r_{BLR}$. Still, this, would require almost unrealistically
  efficient proton acceleration, with $f < 10$. Any hadronic
  activity is expected to be completely absent in the blazar
  spectra if the blazar emission zone is located at distances $r >
  r_{BLR}$ (i.e. at distances where the EC(HDR) process dominates), or
  if $f > 10$.  

\section{Discussion} 

According to the analysis presented in the previous sections, the 
broad-band emission of luminous blazars cannot be easily explained in
the framework of the hadronic scenarios or the SSC model, but instead
is consistent with the EC models. In addition, the
EC(HDR) model is favored over the EC(BLR) variant. This, in turn, fixes
the location of the blazar emission zone at $r \sim 1-10$\,pc from the
nucleus. Note that in the EC(BLR) model, the high energy blazar
emission is produced deeply within the millimeter photosphere and
therefore one should not expect simultaneous high amplitude
$\gamma$-ray and mm-band variability. In contrast, in the EC(HDR)
model the blazar zone extends up to the millimeter photosphere, and
thus mm-band variations accompanying $\gamma$-ray flares are expected,
as in fact observed, e.g., during the 2006 outburst event in 3C\,454.3
(Krichbaum et al.  2008). Hence, systematic monitoring of luminous
blazars at mm/submm wavelengths and GeV photon energies is of primary
interest, allowing to confirm our conclusions.

Regardless of the observational verification, advocating the
EC(HDR) model requires addressing the following questions: (i) why is the
blazar emission zone located at some particular (and, quite 
large: $r \sim 10^5 R_g$ for $10^9 M_{\odot}$ black hole!) distance
from the nucleus? (ii) what is the dominant particle acceleration
process involved, and why is it restricted to the relatively narrow
distance range along the outflow? These questions are in fact related,
since the location of the blazar emission zone is primarily
determined by the energy dissipation processes rather than by some
other factors, such as a presence of seed photons for IC
scattering, i.e. the enhanced radiative cooling of jet electrons.  To
justify this statement, in Figure~3 we have plotted energy densities
of different photon fields (and also of the jet magnetic field), all
as measured in the jet rest frame, as functions of the distance from
the center $r$.  Here we assumed a fully developed outflow with the main
parameters corresponding to the analysis presented in the previous
sections. As shown, the total energy density of different fields
contributing to radiative cooling of jet particles, $u'_{tot} = u'_B +
u'_{BLR}+u'_{HDR}+...$, decreases along the jet roughly as $u'_{tot}
\propto r^{-2}$, being initially dominated by the magnetic field
energy density ($r \lesssim 0.03$\,pc), later by the energy density of
the BLR ($r \sim 0.03-1$\,pc), and next by the energy density of the
HDR ($r \gtrsim 1$\,pc).  Thus, no particular distance scale is
favored as long as solely the $u'_{tot}$ parameter is considered.

The location of the blazar emission zone must be then determined by
the energy dissipation/particle acceleration processes within
relativistic outflow, responsible for generation of the non-thermal
electron population.  In general (though most likely also quite
naively), one can expect that the dominant acceleration mechanism
should be related to the magnetic reconnection process in the case of
a jet dominated by the magnetic field, and to the Fermi-type processes
in the case of a matter-dominated outflow. In this context, we note
that the most recent models for the jet formation indicate that AGN
jets are in fact dominated by the Poynting flux (and are stable
against current-driven Z-pinch instabilities) 
up to at least $r \sim 10^3 R_g$  
(see, e.g., McKinney \& Blandford 2009 and references therein).
At further distances from the jet base, AGN outflows may be considered
as fully developed (i.e., accelerated and collimated to the terminal
values of $\Gamma$ and $\theta_j$ parameters), and likely (though not
necessarily) converted to matter-dominated structures due to the development
of various (mainly kink and pinch) current-driven instabilities. Such
instabilities provide a sink for the jet
magnetic field, and energize jet particles by forming shocks within,
and/or injecting turbulence into the outflow.  

Not much is known about the magnetic reconnection in the relativistic
regime;  in particular the accompanying particle acceleration
processes remain elusive (see Lyubarsky 2005; Lyubarsky \& Liverts
2008; Lyutikov \& Uzdensky 2003).  The results of the analysis
presented in this paper, which suggest that bulk of the blazar
emission is produced far away from the jet base, may thus be taken as
an indication for a low efficiency of the electron acceleration within
the innermost, magnetically-dominated parts of quasar jets. That is to
say, the reconnection-driven acceleration processes may still operate
at distances $r < 10^3 R_g$, but the resulting radiative output does 
not seem to be sufficiently strong to dominate over the emission produced 
further out
from the nuclei of such luminous blazars. In other words, the efficiency of
particle acceleration has to increase significantly at some larger
distance from the core.  And indeed, as mentioned above, initially
suppressed current-driven instabilities are supposed to develop
starting from $r \sim 10^3 R_g$, i.e., $r \sim 0.03$\,pc for a $10^9
M_{\odot}$ black hole, where the BLR dominates the radiative cooling
of the jet electrons. Still, this is about two orders of magnitude
below the location of a blazar emission zone favored by our analysis.

One can propose some {\it ad hoc} explanation for the apparent
discrepancy indicated above, such as, for example, the 
stabilizing role of a velocity shear within relativistic outflow 
(cf. Mizuno et al. 2007), 
precluding formation of shocks or strong turbulence up to the
distances of $\sim 10^4 R_g$ and beyond. The other possible
explanation would be to postulate that the energy dissipation
processes associated with the internal shocks formed due to steepening
of kink instabilities around $10^3 R_g$, are much less efficient in
accelerating jet electrons to ultrarelativistic energies than the
analogous mechanism associated with, e.g., extended reconfinement
shocks formed further out from the nucleus. In fact, a role of
reconfinement shocks in shaping radiative and morphological properties
of blazar jets has been discussed previously by several authors
(Jorstad 2001a, b; Cheung et al. 2007; Sikora et al. 2008; 
Nalewajko \& Sikora 2009; Bromberg \& Levinson 2009). The reason
for such a difference in the acceleration efficiency may be due to the
different turbulence/magnetic field conditions for both kink-driven
and reconfinement shocks: note that the most recent analysis of the
shock acceleration in a relativistic regime clearly shows the
importance of background (plasma) conditions around the shock front in
shaping the spectral properties of non-thermal particles generated
there (e.g., Niemiec \& Ostrowski 2004; Sironi \& Spitkovsky 2009).

But should the dominant particle acceleration process in quasar jets
be undoubtedly identified with shocks? In fact, there are strong
indications for this. Note, first, that the maximum energy of
ultrarelativistic electrons responsible for the electromagnetic
emission of luminous blazars is not high ($\gamma \lesssim
10^4$). Meanwhile, by equating the radiative cooling timescale
$t'_{cool} \sim m_e c / \sigma_T \gamma \, u'_{tot}$ with the electron
acceleration timescale $t'_{acc} \sim f \, \gamma m_e c /e B'$, one
gets the maximum available electron energy $\gamma \sim 10^4$ for the
acceleration efficiency parameter as low as $f^{-1} \sim 10^{-7} -
10^{-8}$ at any distance $0.01$\,pc$\lesssim r \lesssim 10$\,pc. In
other words, the maximum electron energy, if limited solely by the
radiative energy losses, might be expected to be much higher than
observed, since the often invoked `maximum' efficiency of particle
acceleration in cosmic plasma corresponds to $f^{-1} \sim
0.1$ (see also in this context Inoue \& Takahara 1996). 
Second, the electron energy distribution reconstructed from the
observed broad-band electromagnetic spectra of luminous blazars seems
to be best approximated by a broken power-law $N_{\gamma} \propto
\gamma^{-s}$ with $s \leq 2$ for $\gamma < \gamma_{cr} \lesssim 10^3$,
and $s > 2$ for $\gamma > \gamma_{cr}$. This manifests itself in the
so-called `blazar sequence' (Fossati et al. 1998), which, even though
it is still pending the observational confirmation, 
reflects the robust finding 
that the peak electron energy in powerful, quasar-hosted blazars is always
$\gamma_{cr} \lesssim 10^3$ (see Celotti \& Ghisellini 2008). In fact,
exactly this kind of electron spectrum is expected to form at
relativistic (perpendicular) shock mediated by cold protons, for which
the proton inertia (and not the radiative cooling rate!)  determines
the critical electron energy $\gamma_{cr} \lesssim m_p/m_e \sim 10^3$,
as discussed in Stawarz et al. (2007) for the case of terminal shocks
in quasar jets.  

In other words, the electron energy distribution `reconstructed' from
the broad-band emission spectra of luminous blazars indicates that the
appropriate acceleration processes are related to relativistic shocks
formed in a matter (proton) dominated outflow. This assures
self-consistency of the external Compton models favored by our
analysis. We note in this context, that the low-energy segment of the
electron energy distribution discussed above ($\gamma < \gamma_{cr}$)
is expected to form not due to the diffusive (1st-order Fermi) shock
acceleration (which may operate only at $\gamma > \gamma_{cr}$), but
due to interactions of relativistic electrons (with gyroradii smaller
than Larmor radii of cold protons) with turbulence and electromagnetic
waves \emph{within} the shock front. The best studied process of this
type is a resonant absorption of proton cyclotron emission (produced
by cold hadrons reflected from a magnetized perpendicular shock front)
by the jet electrons, as discussed first by Hoshino et
al. (1992). Interestingly, the most recent analysis of this mechanism
(Amato \& Arons 2006) reveals a power-law form of the electron energy
distribution $N_{\gamma} \propto \gamma^{-s}$ with $1 < s \lesssim 2$
(depending on the plasma composition) within the electron energy range
$1 < \gamma < \Gamma_{sh} m_p/m_e$, where $\Gamma_{sh}$ is the shock
bulk Lorentz factor in the upstream plasma rest frame. This is in fact
in a very good agreement with the hard X-ray spectra observed in
luminous blazars, which in turn often exhibit X-ray spectral indices
$\alpha_x < 0.5$ (see Table~1), and with the general form of the
electron spectrum invoked in modeling blazar spectra (Celotti \&
Ghiselini 2008; Ghisellini \& Tavecchio 2008).

\section{Conclusions}

\noindent
$\bullet$ Hadronic models cannot reproduce the very hard X-ray spectra 
observed in a number of
luminous blazars; they also require extremely (almost unrealistically)
efficient acceleration of relativistic protons to the highest energies
within the inner parts of the outflow, and the jet
kinetic power exceeding the Eddington luminosity  by orders of magnitude.

\noindent
$\bullet$ The SSC model, which can account for the hard X-ray spectra
and also for the synchrotron and IC peak frequencies observed in 
luminous blazars, requires significant departures from the minimum 
power condition in order to explain large luminosity ratio of the low-
and high-energy spectral components; furthermore, in a dense radiative
environment of quasar nuclei, energy density of the external radiation
fields (provided inevitably by BLR and HDR) strongly dominates (in a
jet rest frame) over the energy density of the internal jet synchrotron 
photons, and therefore the external-Compton emission is expected to
dominate over the SSC emission.

\noindent
$\bullet$ The EC models -- both EC(BLR) and EC(HDR) variants -- can easily
account for the large-$q$ states of luminous blazars, their hard X-ray
spectra, and the observed ratio of synchrotron and inverse-Compton
peak frequencies; in addition, the main jet parameters implied by
these models (such as $\Gamma$, $L_j$, etc.) are consistent with a
number of different observational constraints.

\noindent
$\bullet$ Lack of the bulk-Compton and the Klein-Nishina features in the 
broad-band spectra of luminous blazars, as well as the absence of the
low-energy cut-off in their X-ray continua, seem to favor the EC(HDR)
model over the EC(BLR) variant; however, these particular constraints have
to be taken with caution, and are not definitive as yet.

\noindent
$\bullet$ The most promising observational discrimination
between the EC(BLR) and EC(HDR) models may be provided by confirmation
of the characteristic variability timescale of luminous blazars being
days/weeks, and by the detection of high-amplitude variability
in mm-band systematically accompanying $\gamma$-ray flares. 
The previous observational results seem to be in fact in agreement with
the predictions of the EC(HDR) model; more high quality, well-sampled data are
however needed in this context, and these are expected to be provided
in a near future by recently launched {\it Fermi} Gamma-ray Space
Telescope surveying the whole sky, 
in conjunction with well-sampled multi-band observing campaigns.  

\noindent
$\bullet$ The electron energy distribution `reconstructed' from the
broad-band emission spectra of luminous blazars indicates that the
appropriate acceleration processes are related to relativistic shocks
formed in a proton-dominated outflow. This assures self-consistency of
the favored EC models, in which the blazar emission zone is located
far away from the nucleus ($r > 10^3 R_g$), where the jet may be
considered as already fully formed (i.e., accelerated and collimated),
and converted to a matter dominated structure.

\acknowledgments

We acknowledge financial support by NASA grants NNX08AZ77G and NNX09AG12G, 
by the Department of Energy contract to SLAC no. DE-AE3-76SF00515,
by the Polish MNiSW grant N N203 301635, N N203 380336, and the Polish 
Astroparticle Network 621/E-78/BWSN-0068/2008.

\begin{deluxetable}{cccccc}
\tabletypesize{\scriptsize}
\tablecaption{Luminous Blazar Sources with the Hardest Recorded X-ray Spectra}
\tablewidth{0pt}
\tablehead{
\colhead{name}
& \colhead{$z$}
& \colhead{$\alpha_x$}
& \colhead{$\alpha_{\gamma}^E$}
& \colhead{$\alpha_{\gamma}^F$}
& \colhead{Ref}\\
(1) & (2) & (3) & (4) & (5) & (6)
}
\startdata
S5~0212+73         & $2.367$ & $0.32\pm 0.19$ & ---            & ---             & Sambruna et al. (2007) \\
PKS~0229+13        & $2.059$ & $0.39\pm 0.09$ & ---            & ---             & Marshall et al. (2005) \\
PKS~0413-21        & $0.808$ & $0.39\pm 0.12$ & ---            & ---             & Marshall et al. (2005) \\
PKS~0528+134       & $2.060$ & $0.12\pm 0.26$ & $1.46\pm 0.04$ & $1.54 \pm 0.09$ & Donato et al. (2005) \\
PKS~0537-286       & $3.104$ & $0.27\pm 0.02$ & $1.47\pm 0.60$ & ---             & Reeves et al. (2001) \\
PKS~0745+241       & $0.409$ & $0.35\pm 0.12$ & ---            & ---             & Marshall et al. (2005) \\
SWIFT~J0746.3+2548 & $2.979$ & $0.17\pm 0.01$ & ---            & ---             & Watanabe et al. (2009) \\
PKS~0805-07        & $1.837$ & $0.20\pm 0.20$ & $1.34\pm 0.29(?)$ & ---          & Giommi et al. (2007) \\
S5~0836+710        & $2.172$ & $0.34\pm 0.04$ & $1.62\pm 0.16$ & ---             & Donato et al. (2005) \\
RGB J0909+039      & $3.200$ & $0.26\pm 0.12$ & ---            & ---             & Giommi et al. (2002) \\
PKS~1127-145       & $1.184$ & $0.20\pm 0.03$ & $1.70\pm 0.31$ & $1.69 \pm 0.18$ & Siemiginowska et al. (2008) \\
PKS~1424-41        & $1.522$ & $0.20\pm 0.30$ & $1.13\pm 0.21$ & ---             & Giommi et al. (2007) \\
GB~1428+4217       & $4.715$ & $0.29\pm 0.05$ & ---            & ---             & Fabian et al. (1998) \\
PKS~1510-089       & $0.360$ & $0.23\pm 0.01$ & $1.47\pm 0.21$ & $1.48 \pm 0.05$ & Kataoka et al. (2008) \\
PKS~1830-211       & $2.507$ & $0.09\pm 0.05$ & $1.59\pm 0.13$ & ---             & De Rosa et al. (2005) \\
PKS~2149-306       & $2.345$ & $0.38\pm 0.08$ & ---            & ---             & Donato et al. (2005) \\
PKS~2223+210       & $1.959$ & $0.31\pm 0.26$ & ---            & ---             & Donato et al. (2005) \\
3C~454.3           & $0.859$ & $0.34\pm 0.06$ & $1.21\pm 0.06$ & $1.41 \pm 0.02$ & Donato et al. (2005) \\
\enddata
\tablecomments{
(1) Name of a source;
(2) Redshift of a source $z$;
(3) X-ray spectral index $\alpha_x$;
(4) \emph{EGRET} $\gamma$-ray spectral index $\alpha_{\gamma}^E$ (Hartman et al. 1999);
(5) \emph{FERMI} $\gamma$-ray spectral index $\alpha_{\gamma}^F$ (Abdo et al. 2009b);
(6) References.}
\end{deluxetable}

\begin{figure}
\begin{center}
\includegraphics[width=0.9\textwidth,bb = 55 175 530 660]{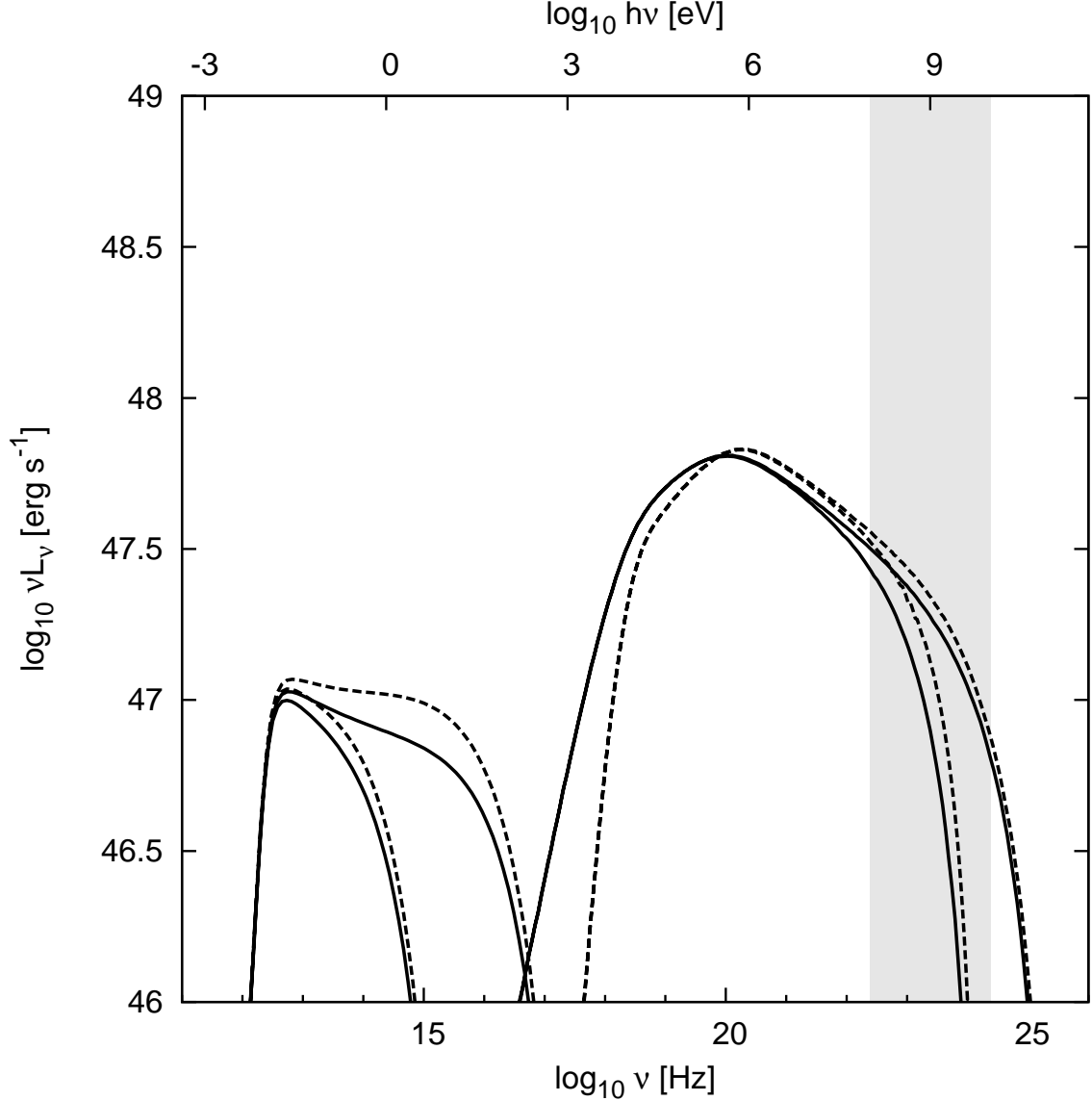}
\end{center}
\caption{Broadband spectra of fiducial blazars showing the Klein-Nishina 
effects for the EC(BLR) model. Solid lines are calculated
  for a power-law type external radiation with photon index $\alpha=0$,
  while dashed lines show models with monochromatic external radiation
  (approximated by black-body type emission). Two families of models
  are shown, corresponding to the injection of the electron energy distribution
  $N_{\gamma} \propto \gamma^{-2.4}$ with maximal electron energy $\gamma_{max} 
  = 10^3$, and $10^4$ (lower and upper curves, respectively). All models are 
  calculated for $q=10$. Shaded area indicates the energy range observable 
by the {\it Fermi}/LAT.}
\label{fig:kn10}
\end{figure}

\begin{figure}
\begin{center}
\includegraphics[width=0.9\textwidth,bb = 55 175 530 660]{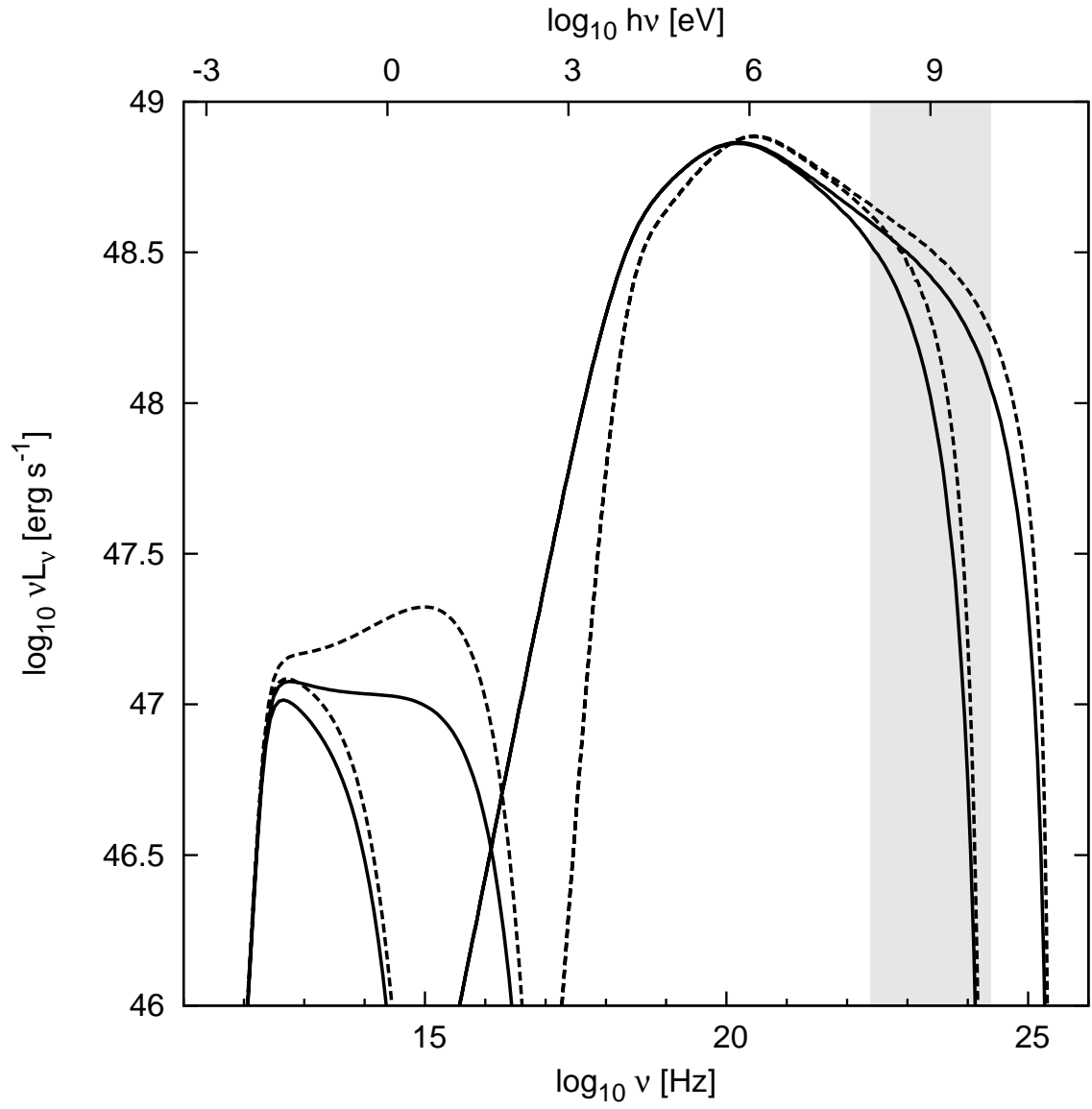}
\end{center}
\caption{Same as Figure~\ref{fig:kn10}, but with $q=100$.}
\label{fig:kn100}
\end{figure}

\begin{figure}[ht]
\includegraphics[width=\textwidth]{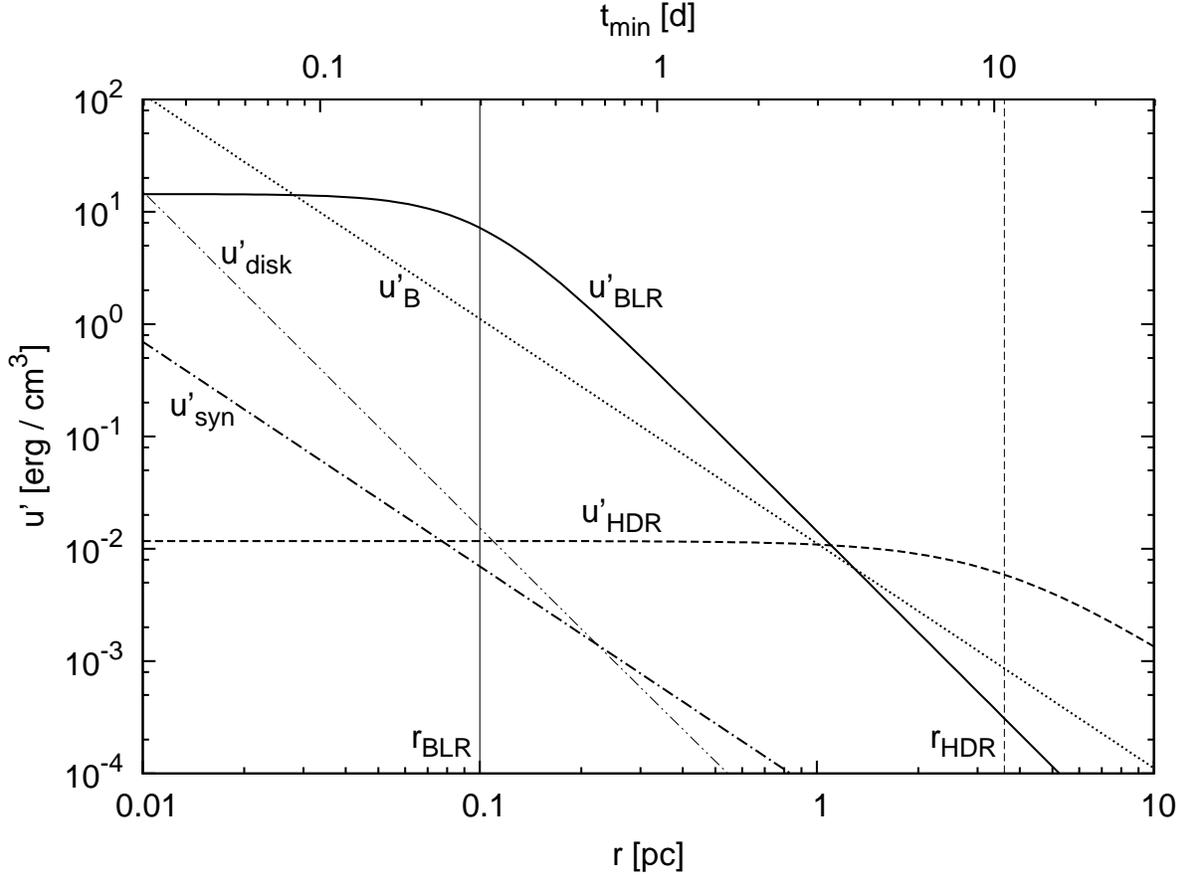}
\caption{Energy densities of broad line emission (\emph{solid line}), hot dust radiation (\emph{dashed line}), magnetic field (\emph{dotted line}), synchrotron radiation (\emph{dot-dashed line}) and accretion disk emission (\emph{thin double-dot-dashed line}) as seen in the jet comoving frame, as a function of distance from the central engine. 
The shortest timescale of flares corresponding to the jet radius $R=r/\Gamma$ at a given distance from the center $r$, as measured in the observer frame, is shown on the \emph{top axis}. The characteristic radii of broad line region $r_{BLR}$ and dusty torus $r_{HDR}$ are indicated with a \emph{solid and dashed vertical line}, respectively. We assumed here: jet bulk Lorentz factor $\Gamma=20$, accretion disk luminosity $L_{disk}=10^{46}\;{\rm erg/s}$, synchrotron luminosity $L_{syn}=10^{47}\;{\rm erg/s}$, magnetic flux $L_B=10^{46}\;{\rm erg/s}$, broad line region covering factor $\xi_{BLR}=0.1$, dusty torus covering factor $\xi_{HDR}=0.1$ and dust temperature $T_{dust}=10^3\;{\rm K}$. We assume that beyond $r_{BLR}$ stratification of the broad line emission takes form $dL_{BLR}/d\ln r\propto 1/r$, and hence $u'_{BLR}\propto r^{-3}$ for $r > r_{BLR}$. Radiation energy density from the accretion disk is calculated using formula 
$u_d' \simeq 0.28 \, (L_{disk}/4\pi r^2 c)\,(R_g/r) \Gamma^2$ for $R_g$ of the fast rotating black hole with mass $M_{BH}=10^9 M_{\odot}$ (Dermer \& Schlickeiser 2002).}
\end{figure}

\end{document}